\documentclass{article}
\usepackage{spconfa4,amsmath,graphicx}

\newcommand{\Fig}[1]{\textbf{Fig. \ref{fig:#1}}} % refer to figure
\newcommand{\Table}[1]{\textbf{Table \ref{tb:#1}}} % refer to table
\newcommand{\Eq}[1]{Eq. (\ref{eq:#1})} % refer to equation
\newcommand{\Sec}[1]{\textbf{　Section \ref{sec:#1}}} % refer to section
\newcommand{\Figure}[3]{\vspace{-1mm} \includegraphics[width=#1,clip]{#2.eps} \vspace{-3mm} \caption{#3} \vspace{-5mm} \label{fig:#2}} 

\renewcommand{\Vec}[1]{\textrm{\boldmath $#1$}} % Vector
\newcommand{\Section}[1]{\vspace{-2mm} \section{#1} \vspace{-1mm}}
\newcommand{\Subsection}[1]{\vspace{-2mm} \subsection{#1} \vspace{-1mm}}

\newcommand{\pt}[1]{\left(#1\right)} % ()
\newcommand{\br}[1]{\left[#1\right]} % []
\newcommand{\x}{ \Vec{x} } % Vector x
\newcommand{\y}{ \Vec{y} } % Vector y
\newcommand{\haty}{ \Vec{\hat y} } % Vector y
\newcommand{\drawfig}[4]{ % draw figure 
  \begin{figure}[#1]
  \begin{center}
  \Figure{#2}{#3}{#4} 
  \end{center} 
  \end{figure}
}
\title{Phase reconstruction from amplitude spectrograms\\
based on von-Mises-distribution deep neural network}

\name{Shinnosuke Takamichi$^\dagger$, Yuki Saito$^\dagger$, Norihiro Takamune$^\dagger$, Daichi Kitamura$^\ddagger$, and Hiroshi Saruwatari$^\dagger$}
\address{$^\dagger$ Graduate School of Information Science and Technology, The University of Tokyo, Japan. \\
$^\ddagger$ Department of Electrical and Computer Engineering, National Institute of Technology, \\Kagawa College, Japan.
}

\begin{document}
\ninept
\maketitle
\setlength{\abovedisplayskip}{3pt}
\setlength{\belowdisplayskip}{3pt}

\begin{abstract}
This paper presents a deep neural network (DNN)-based phase reconstruction from amplitude spectrograms. In audio signal and speech processing, the amplitude spectrogram is often used for processing, and the corresponding phase spectrogram is reconstructed from the amplitude spectrogram on the basis of the Griffin-Lim method. However, the Griffin-Lim method causes unnatural artifacts in synthetic speech. Addressing this problem, we introduce the von-Mises-distribution DNN for phase reconstruction. The DNN is a generative model having the von Mises distribution that can model distributions of a periodic variable such as a phase, and the model parameters of the DNN are estimated on the basis of the maximum likelihood criterion. Furthermore, we propose a group-delay loss for DNN training to make the predicted group delay close to a natural group delay. The experimental results demonstrate that 1) the trained DNN can predict group delay accurately more than phases themselves, and 2) our phase reconstruction methods achieve better speech quality than the conventional Griffin-Lim method.
\end{abstract}


\Section{Introduction} \label{sec:intro} \vspace{-1mm}
A variety of audio signal processing and machine learning-based methods, such as audio source separation and speech enhancement, involve processing the amplitude spectrogram obtained via short-term Fourier transform (STFT). Also, statistical parametric text-to-speech synthesis \cite{zen09} is shifting from vocoder-based (source filter model-based) to STFT-based frameworks \cite{takaki17fft,wang17tacotron,saito18fftssgan}. To produce synthetic speech, we require the corresponding phase spectrogram, but it is often unavailable. The Griffin-Lim method \cite{griffin84phasereconstruction} is a well-known example that iteratively estimates of the phase spectrogram through STFT and inverse STFT while fixing the amplitude spectrogram. This signal-processing-based method has high portability without a priori training but causes unnatural artifacts in synthetic speech. Therefore, this paper proposes building a \textit{trainable} phase reconstruction method using generative models.

A deep neural network (DNN) is a powerful generative model. There are two types of distributions: non-parametric \cite{goodfellow14gan,saito18advss,li15momentmatchingnetwork,takamichi17moment} and parametric ones. This paper addresses the latter. There are many types of the parametric approaches related to the Gaussian distribution, e.g., isotropic multivariate Gaussian \cite{zen13dnn} and temporal-delta-constrained Gaussian \cite{wu15dnnmge,hashimoto15} (a.k.a., trajectory DNN in statistical parametric speech synthesis). The loss function (e.g., mean squared error) for training DNNs is often derived to minimize the negative log-likelihood of the distribution. The straightforward way to predict a phase spectrogram from an amplitude spectrogram is to use these models. However, the Gaussian distribution is inappropriate for modeling distributions of a phase that is a periodic variable with a period of $2\pi$.

This paper proposes phase reconstruction from amplitude spectrograms based on the \textit{von-Mises-distribution DNN}. The von Mises distribution is a probability distribution on the circle \cite{mardia99statistics}, which is suitable for modeling periodic variables. The von-Mises-distribution DNN is a generative model that has the von Mises distribution as a conditional probability distribution. The von-Mises-distribution shallow neural network was originally proposed by Nabney et al. \cite{nabney95vonmisesnn}, and this paper utilizes it for predicting the phase spectrogram from the amplitude spectrogram. The loss function for DNN training, named the \textit{phase loss} (see \Fig{eps/overview}), is defined by minimizing likelihoods of the von Mises distribution. Also, we propose another loss function named \textit{group-delay loss}, which has a stronger connection to the amplitude spectrum \cite{itaku87asrgroupdelay}. The group-delay loss is used to make group delay of the predicted phase close to that of the target phase. Since the group delay and group-delay loss are differentiable by the phase, the DNN can be trained by a standard backpropagation algorithm. We conduct objective and subjective evaluations to evaluate the effectiveness of the proposed methods. The results demonstrate that 1) the trained DNN can predict group delay accurately more than phases themselves, and 2) our phase reconstruction method outperforms the conventional Griffin-Lim method in terms of quality of synthesized speech.

	\drawfig{t}{0.85\linewidth}{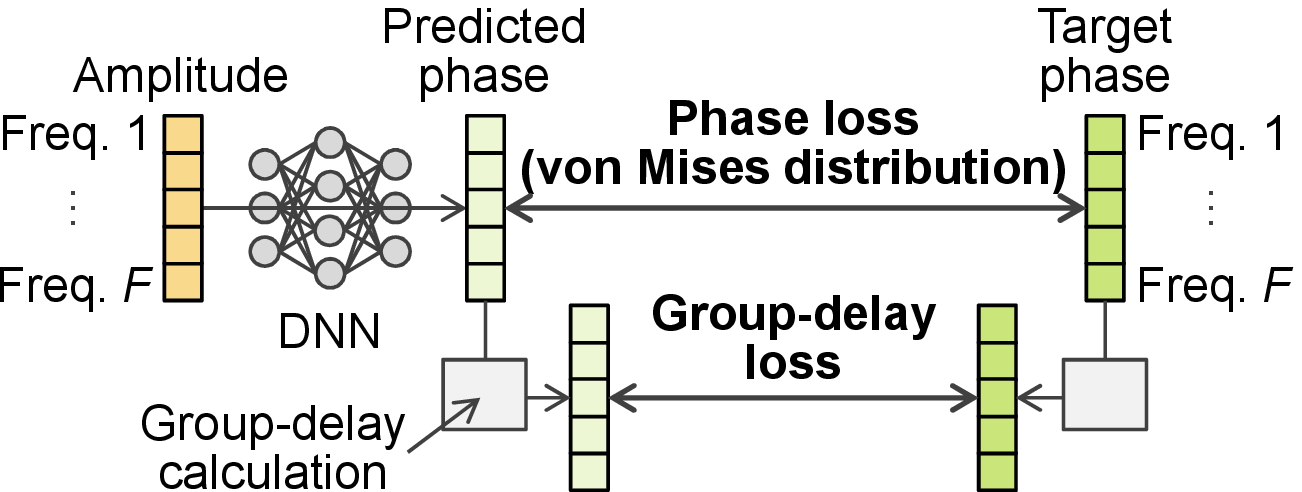}
    {Overview of proposed phase reconstruction method. This figure shows frame-by-frame phase prediction rather than multi-frame or sequence-wise prediction for clear illustration.}
    
\Section{Griffin-Lim phase reconstruction} \vspace{-1mm}
This section briefly describes the conventional Griffin-Lim method \cite{griffin84phasereconstruction}. The Griffin-Lim method is a signal-processing-based iterative algorithm to reconstruct a phase spectrogram from the amplitude spectrogram. Let $\x = \br{\x_1, \cdots, \x_t, \cdots, \x_T}$ and $\y = \br{\y_1, \cdots, \y_t, \cdots, \y_T}$ be amplitude and phase spectrograms, respectively.
$\x_t = \br{x_{t,0}, \cdots, x_{t,f}, \cdots, x_{t,F}}$ and $\y_t = \br{y_{t,0}, \cdots, y_{t,f}, \cdots, y_{t,F}}$ are the amplitude and phase at frame $t$, respectively. $f$ is the frequency index, and $F$ corresponds to the Nyquist frequency. Both $x_{t,f}$ and $y_{t,f}$ are real-valued variables, but only $y_{t,f}$ is a variable with a period of $2\pi$. The Griffin-Lim method randomly initializes $\y$ first. 1) Then it takes inverse STFT to obtain a waveform from $\x$ and $\y$. 2) The method takes STFT to re-obtain $\y$ from the waveform. 3) We substitute the original $\x$ for the re-obtained $\x$ and then go back to step 1). These inverse STFT and STFT are iteratively performed until they have converged. The method can reconstruct the phase spectrogram consistent with the given amplitude spectrogram but makes some artifacts in the synthesized speech, e.g., extra reverberation and phasiness owing to inappropriate initialization of $\y$.

\Section{Phase reconstruction based on von-Mises-distribution DNNs}
This section introduces the von-Mises-distribution DNN and proposes loss functions for the DNN training.

	\Subsection{von Mises distribution}
	The von Mises distribution $P^{(\rm vM)} \pt{\cdot}$ \cite{mardia99statistics} is a probability distribution for a periodic variable $y_{t,f}$,
    given as
        \begin{align}
        P^{(\rm vM)} \pt{y_{t,f}; \mu, \kappa} = 
        	\frac{\exp\pt{\kappa \cos \pt{y_{t,f} - \mu} }}
            {2 \pi I_0 \pt{\kappa}},
        \end{align}
	where $\mu$ is the mean direction of the distribution (analogous to the mean of the Gaussian distribution), 
    $\kappa$ is the shape parameter (analogous to the precision of the Gaussian distribution), and 
    $I_0\pt{\cdot}$ is the modified Bessel function of the first kind of order 0.
	The negative log likelihood given $\y_t$ is:
        \begin{align}
        	- \log { P^{(\rm vM)} \pt{\y_t; \mu, \kappa} } \propto
            - \sum\limits_{f=0}^{F} \cos \pt{ y_{t,f} - \mu } + \textrm{Const.}, \label{eq:nll}
        \end{align}    
	where $\textrm{Const.}$ is a value constant to $\mu$.
	Not only $y_{t,f}$ but also a maximum likelihood estimate of $\mu$ has a period of $2\pi$.

	\Subsection{DNN training}
    We train a DNN that has the von Mises distribution as a conditional probability distribution.
    The mean direction is predicted from $\x$ at each frame and frequency. 
    Here, let $\Vec{G}\pt{\cdot}$ be the DNN. 
    The predicted phase (mean direction) $\haty = \br{\haty_1, \cdots, \haty_t, \cdots, \haty_T}$ is given as $\haty = \Vec{G}\pt{\x}$.
    The following sections propose two loss functions for estimating model parameters of $\Vec{G}\pt{\cdot}$: 
    phase loss $L_{\rm ph}\pt{\y_t, \haty_t}$ and group-delay loss $L_{\rm gd}\pt{\y_t, \haty_t}$.

		\vspace{-1mm}
		\subsubsection{Phase loss} \label{sec:phase_loss} \vspace{-1mm}
        The phase loss function is derived from \Eq{nll} as follows:
            \begin{align}
            L_{\rm ph}\pt{\y_t, \haty_t} = \sum\limits_{f = 0}^{F} - \cos\pt{ y_{t,f} - {\hat y}_{t,f}}. \label{eq:phase_loss}
            \end{align}
		Model parameters of $\Vec{G}\pt{\cdot}$ are iteratively updated by backpropagation algorithm to minimize this loss function.
		The minimum point of ${\hat y}_{t,f}$ is periodic, i.e., ${\hat y}_{t,f} = y_{t,f} \pm 2 \pi N$,
        where $N$ is an arbitrary integer value.

		\vspace{-1mm}
		\subsubsection{Group-delay loss} \vspace{-1mm}
        Group delay of speech has a high potential in speech processing,
        such as speech and 
        speaker recognition \cite{itaku87asrgroupdelay,padmanabhan09speakerrecognitiongroupdelay}.
        The group delay is defined as the negative derivative of phase by frequency.
        In general, speech production via a human vocal tract is well modeled
        as an all-pole filter defined by $A(\omega)\exp(j\phi (\omega))$, where
        $A(\omega)$ and $\phi (\omega)$ are the amplitude and phase functions, respectively,
		and $\omega = \pi f / F$ is the angular frequency.
        If the filter has $P$ complex-valued poles $z_p = r_p \exp (j\omega_p)$ ($p=1,..., P$),
        we have $\log A(\omega) = \log \prod_{p=1}^{P} A_p (\omega)
        = 2 \sum_{p=1}^{P} \sum_{n=1}^{\infty} r_p^n / n \cos n(\omega -\omega_p)$,
        where $A_p (\omega)$ is the amplitude of the single-pole model
        for the specific $p$-th pole \cite{itaku87asrgroupdelay}.
        Then, the group delay can be given by
            \begin{align}
            -\frac{d \phi (\omega)}{d \omega}
            = c \sum_{p=1}^P \sum_{n=1}^{\infty} n \cos n (\omega - \omega_p) \int_{-\pi}^{\pi} \log A_p(\omega) \cos (n\omega) d\omega,
            \end{align}
        where $c$ is a constant value; this shows the strong correlation between the group delay and the amplitude spectrum, which 
        motivates us to utilize the group delay as a \textit{regularization term}.

        %The group delay is defined as the derivative of $\y$ by frequency, and 
        In this paper, we approximate the group delay at frame $t$ and frequency bin $f$ with the following equation:
            \begin{align}
            \Delta y_{t,f} = -\pt{y_{t,f+1} - y_{t,f}}. \label{eq:group_delay}
            \end{align}
        Since $\Delta y_{t,f}$ is also a periodic variable, 
        the group-delay loss is defined as similar to \Eq{phase_loss}:
          \begin{align}
          L_{\rm gd}\pt{\y_t, \haty_t} = \sum\limits_{f = 0}^{F} - \cos\pt{ \Delta y_{t,f} - \Delta {\hat y}_{t,f} }. \label{eq:group_delay_loss}
          \end{align}
        The group-delay loss makes $\Delta \hat y_{t,f}$ close to $\Delta y_{t,f}$.
        Because \Eq{group_delay} for $\haty_t$ is a linear transformation of $\y_t$, 
        the backpropagation algorithm can be used as the same as in \Sec{phase_loss}.

		\vspace{-1mm}
        \subsubsection{Multi-task learning} \vspace{-1mm}
        On the basis of the multi-task learning formulation, the DNN can be trained using both phase loss and group-delay loss.
        The loss function $L \pt{\y_t, \haty_t}$ is:
            \begin{align}
            L \pt{\y_t, \haty_t} = L_{\rm ph}\pt{\y_t, \haty_t} + \alpha L_{\rm gd}\pt{\y_t, \haty_t}, \label{eq:multi_task}
            \end{align}    
        where $\alpha$ is the weight of the secondary task (group-delay loss).
        Note that, since ranges of \Eq{phase_loss} and \Eq{group_delay_loss} are the same,
        no scale normalization factor is required.
    
	\Subsection{Discussion}
      \drawfig{t}{0.85\linewidth}{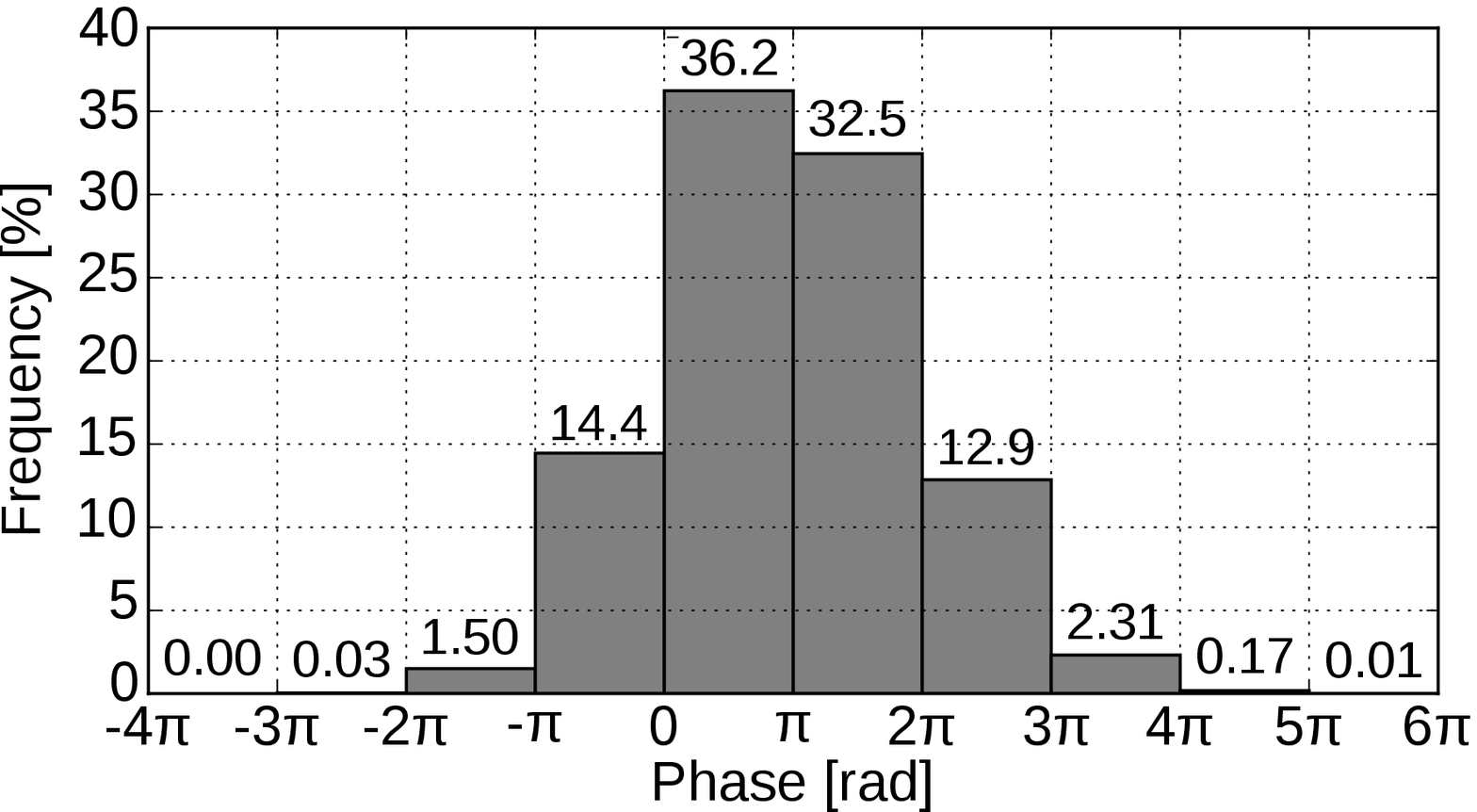}
          {Histogram of predicted phases. The target phases have a range of $\br{0, 2\pi}$, but the predicted phases have a range of $\br{-4\pi, 6\pi}$. 
          All frames and frequency bins of the evaluation data in \Sec{setup} are represented in this figure.
          }
    
    % generalization
	The generalized version of the von Mises distribution is the \textit{generalized cardioid distribution} \cite{jones05generalizedcardioid} given as
      \begin{align}
      P^{(\rm GC)} \pt{y_{t,f}; \mu, \kappa, \psi} &= \nonumber \\
      	& \hspace{-2cm} \frac{ \pt{\cosh \pt{\kappa\psi}}^{1/\psi} \pt{1+\tanh\pt{\kappa\psi}\cos\pt{y_{t,f} - \mu}}^{1/\psi} }
        {2 \pi P_{1/\psi} \pt{ \cosh\pt{\kappa\psi} }},
      \end{align}
	where $P_{1/\psi}$ is the associated Legendre function of the first kind of degree $1/\psi$ and order 0.
    The von Mises distribution is the special case ($\psi \rightarrow 0$) of this distribution.
	Also, this distribution is equivalent to the cardioid distribution and the wrapped Cauchy distribution for $\psi = 1$ and $\psi = -1$, respectively.
    The negative log likelihood for $\mu$ for the cardioid and wrapped Cauchy distributions are equal to \Eq{nll}.
    Therefore, DNNs with these two distributions are trained in the same manner.
    One possible way to extend our work is to model phases using this generalized cardioid distribution.
    Other possible ways are to use an asymmetric distribution \cite{abe11sinecircular} and mixture model.
    
	% unwrapping
    As described in \Sec{phase_loss}, the phase loss is minimized when ${\hat y}_{t,f} = y_{t,f} \pm 2 \pi N$ for arbitrary $N$.
    Therefore, von-Mises-distribution DNNs suffer from an exploding value of ${\hat y}_{t,f}$.
    To investigate this, \Fig{eps/hist_phase} plots a histogram of the predicted phase spectrograms.
    We can see that the predicted phase spectrograms have a wider range than the target phase ($\br{0, 2\pi}$),
    but the value does not explode.

\Section{Experimental evaluation}
	\Subsection{Experimental setup} \label{sec:setup}
    Evaluations were performed using the JSUT corpus \cite{jsut}, a free Japanese speech corpus uttered by a female speaker.
    We used 5,000 utterances (approx. 6 hours) of the subset BASIC5000 for training and 300 utterances of the subset ONOMATOPEE300 for evaluation.
    Speech signals were downsampled at a rate of 16~kHz.
    The window length, shift length, and FFT length were set to 400 samples (25~ms), 80 samples (5~ms), and 512 samples, respectively.
    The Hamming window was used.
    Features fed to a DNN were the joint vectors of the log amplitude spectra at current and $\pm 2$ frames, and they were normalized to have zero-mean unit-variance.
    The DNN architectures were Feed-Forward networks that included $3\times 1024$-unit gated linear unit \cite{dauphin16gatedlinear} hidden layers.
    The activation of the output layer was a linear function.
    We empirically explored DNN architectures within Feed-Forward networks and found that the gated linear unit is significantly better than a rectified linear unit (ReLU) \cite{glorot11relu}
    or LeakyReLU \cite{maas13leakyrelu} hidden units.
    The DNN parameters were randomly initialized and the AdaGrad algorithm \cite{duchi11adagrad} with its learning rate set to 0.001 was used for the optimization algorithm.

	We compared the conventional Griffin-Lim and three proposed phase reconstruction methods.
    	\begin{itemize} \leftskip 1.0cm \itemsep -0mm \vspace{-1mm}
    	\item [\textbf{PH}] only phase loss (\Eq{phase_loss})
        \item [\textbf{GD}] only group-delay loss (\Eq{group_delay_loss})
        \item [\textbf{PH+GD}] multi-task learning (\Eq{multi_task})
    	\end{itemize}  \vspace{-1mm}
    In the Griffin-Lim method, phase spectrograms were randomly initialized. The number of iterations was set to 100.
    The weight $\alpha$ for multi-task learning was set to $0.1$.
    In the proposed methods, phases in the low frequency band are first estimated by DNNs, and 
    those in the remaining frequency bins are randomly generated.
    We used three settings of frequency bands of the predicted phase spectrograms: 0-2 (96 dim.), 0-4 (128 dim.), and 0-8 kHz (257 dim.).
    After predicting the phase spectrogram by DNNs, we further applied the Griffin-Lim method to refine the phase.
	The number of iterations of the phase refinement was 100.

	\Subsection{Prediction accuracy}
        \drawfig{t}{0.85\linewidth}{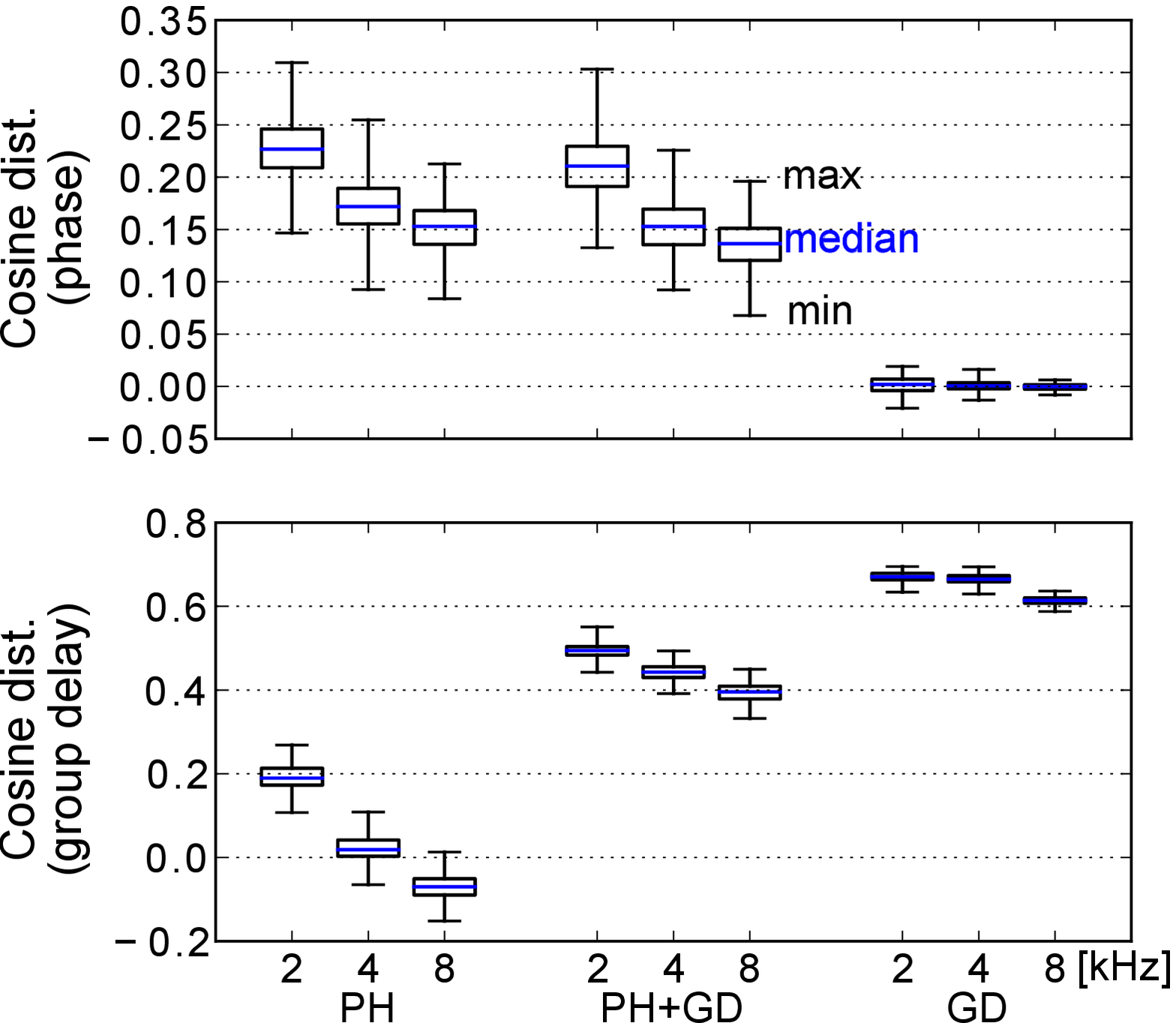}
        {Box plots of cosine distances between target and predicted phases (upper) and group delays (lower). 
        The box indicates the first and third quartiles.}  

        \begin{table}[t]
        \centering
        \caption{Results of preference tests: conventional Griffin-Lim method vs. proposed methods.
        \textbf{Bold} indicates preferred method that has a $p$-value smaller than $0.05$ }
        \label{tb:subj_conv_vs_proposed}
        \begin{tabular}{|r|cc|l|}
        \hline Method A & Scores & $p$-value & Method B \\
        \hline
        \hline Griffin-Lim & 0.497 vs. 0.503 & 0.871 & PH (2 kHz) \\
        \hline Griffin-Lim & 0.280 vs. \textbf{0.720} & $< 10^{-9}$ & PH (4 kHz) \\
        \hline Griffin-Lim & 0.277 vs. \textbf{0.723} & $< 10^{-9}$ & PH (8 kHz) \\

        \hline Griffin-Lim & 0.453 vs. \textbf{0.547} & 0.022 & PH+GD (2 kHz) \\
        \hline Griffin-Lim & 0.233 vs. \textbf{0.767} & $< 10^{-9}$ & PH+GD (4 kHz) \\
        \hline Griffin-Lim & 0.247 vs. \textbf{0.753} & $< 10^{-9}$ & PH+GD (8 kHz) \\

        \hline Griffin-Lim & 0.447 vs. \textbf{0.553} & 0.009 & GD (2 kHz) \\
        \hline Griffin-Lim & 0.463 vs. 0.537 & 0.073 & GD (4 kHz) \\
        \hline Griffin-Lim & 0.490 vs. 0.510 & 0.619 & GD (8 kHz) \\     
        \hline
        \end{tabular}    	
        \end{table}  
        
    We evaluated prediction accuracies of phases and group delay. 
    \Fig{eps/obj4} shows cosine distances between target and predicted phases and group delays.
    The distances were averaged over all frames and frequency bins of the predicted phases (0-2, 0-4, or 0-8 kHz).

    The prediction accuracy of ``PH (2 kHz)'' ranges from $0.15$ to $0.31$, and the distribution seems symmetric.
    Also, the accuracy becomes smaller as the frequency band widens (``PH (4, 8 kHz)'').
    This result is natural because phases at higher frequency bins are easily changed by the temporal position of frame analysis.
    On the other hand, when using only group-delay loss (``GD''), 
    group delay can be predicted more accurately than phase for all settings of frequency bands.
    This result shows us that the Feed-Forward DNN can predict group delay better than phases themselves.
    Finally, combined phase loss and group-delay loss (``PH+GD'') predicts phase more accurately than ``GD'' and
    group delay more accurately than ``PH'' for all settings of frequency bands.
    Therefore, we can demonstrate the effectiveness of the proposed multi-task training.

	\Subsection{Comparison of Griffin-Lim and proposed methods}
    To evaluate the effectiveness of the proposed methods, we compared the quality of synthetic speech of Griffin-Lim and proposed methods.
    Preference AB tests (listening tests) on speech quality were performed in our crowdsourcing evaluation system.
    30 listeners participated in each test. Approximately $\$0.46$ were paid to each listener.
    Each listener preferred better-quality speech and answered for ten pairs of samples, i.e.,
    300 answers were obtained in each evaluation.
    Speech samples of pairs of methods were randomly presented to the listeners.
    These settings are also used not only here but also in the following sections.

    \Table{subj_conv_vs_proposed} lists the results. 
    The proposed methods outperform the conventional Griffin-Lim method in all settings of loss functions and frequency bands.
    In particular, ``PH+GD'' always has significantly better scores than the conventional Griffin-Lim method.
    These results demonstrate the effectiveness of the proposed methods.

	\Subsection{Effect of phase refinements}
        \drawfig{t}{0.85\linewidth}{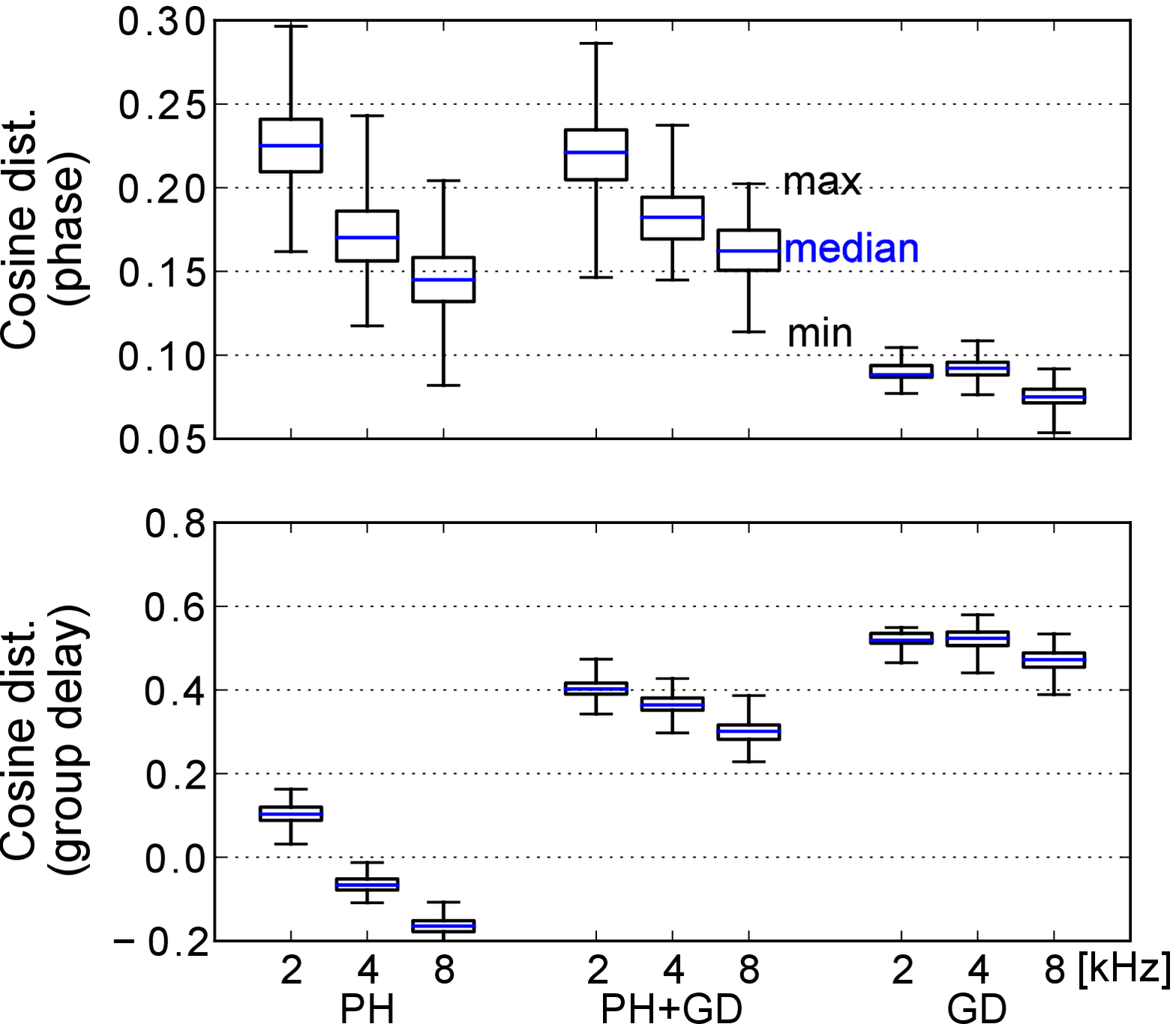}
        {Box plots of cosine distances between predicted and refined phases (upper) and group delays (lower). 
        The box indicates the first and third quartiles.}

        \begin{table}[t]
        \centering
        \caption{Results of preference tests: proposed methods with different frequency bands.
        \textbf{Bold} indicates preferred method that has a $p$-value smaller than $0.05$ }
        \label{tb:subj_freq_band}
        \begin{tabular}{|r|cc|l|}
        \hline Method A & Scores & $p$-value & Method B \\
        \hline        
        \hline PH (2 kHz) & 0.270 vs. \textbf{0.730} & $<10^{-9}$ & PH (4 kHz) \\
        \hline PH (4 kHz) & 0.507 vs. 0.493 & 0.744 & PH (8 kHz) \\

        \hline PH+GD (2 kHz) & 0.223 vs. \textbf{0.777} & $<10^{-9}$ & PH+GD (4 kHz) \\
        \hline PH+GD (4 kHz) & 0.493 vs. 0.507 & 0.744 & PH+GD (8 kHz) \\

        \hline GD (2 kHz) & 0.513 vs. 0.487 & 0.514 & GD (4 kHz) \\
        \hline GD (4 kHz) & \textbf{0.567} vs. 0.433 & 0.001 & GD (8 kHz) \\     
        \hline
        \end{tabular}
        \end{table}

    As explained in \Sec{setup}, after phases were predicted by DNNs, they were refined by the Griffin-Lim method.
	Here, we evaluated effects of the phase refinements. 
    \Fig{eps/refined-phase} shows cosine distances between predicted and refined phases and group delays.
	Tendencies were almost the same as those in \Fig{eps/obj4}.
    When phase loss is used in training (``PH'' and ``PH+GD''), phase information is comparably preserved.
    Similarly, when group-delay loss is used in training (``GD'' and ``PH+GD''), the group-delay information is comparably preserved.
    In the preliminary evaluation, we compared speech quality of refined and unrefined phases (i.e., phases predicted by DNNs were directly used for finally synthesized speech).
    The results demonstrated that unrefined phases had significantly worse speech quality than refined phases.
        
%     We investigated perceptual effects of phase refinements.
%     If there is no differences in quality between synthetic speech with and without phase refinements,
%     simplified and non-iterative waveform synthesis can be used.

%     \Table{subj_refine} lists the results.
%     ``(unrefined)'' in this table means that the predicted phase spectrograms were directly used for waveform synthesis
%     without iterative phase refinements.
%     We can see that speech quality of the unrefined phase is significantly worse than the refined ones
%     even using any loss functions of the proposed methods, and the refined ones were almost always preferred. 
%     Therefore, we need the phase refinements.        

%         \begin{table}[t]
%         \centering
%         \caption{Results of preference tests: effects of phase refinements.
%         \textbf{Bold} indicates a method that the $p$-values is smaller than $0.05$.
%         Phase spectrograms of 0-8 kHz were predicted.}
%         \label{tb:subj_refine}
%         \begin{tabular}{|r|cc|l|}
%         \hline Method A & Scores & $p$-value & Method B \\
%         \hline   
%         \hline PH (unrefined) & 0.038 vs. \textbf{0.962} & $< 10^{-9}$ & PH \\
%         \hline PH+GD (unrefined) & 0.028 vs. \textbf{0.972} & $< 10^{-9}$ & PH+GD \\      
%         \hline GD (unrefined) & 0.014 vs. \textbf{0.986} & $< 10^{-9}$ & GD \\       
%         \hline
%         \end{tabular}
%         \end{table}

	\Subsection{Effect of frequency bands}
    We compared effects of frequency bands of predicted phases within one loss function (``PH'', ``PH+GD'', or ``GD'').
    \Table{subj_freq_band} shows the results of preference AB tests on speech quality.
    In ``PH'' and ``PH+GD,'' speech quality for 0-4 kHz frequency bands is significantly better than that for 0-2 kHz,
    and is comparable with that for 0-8 kHz.
    These results suggest that at least 0-4 kHz frequency bands needs to be predicted but 
    the spectrograms in the higher frequency bands may be excited by random phases
    (this tendency is similar to the harmonics plus noise model \cite{stylianou01harmonicsplusnoise}).
    A curious point is that ``GD'' has different tendencies:
    speech quality for 0-8 kHz frequency bands was significantly worse than that for 0-4 kHz.
    We will investigate the reason for this.

	\Subsection{Effect of group-delay loss}
    We investigated the effectiveness of group-delay loss compared with phase loss.
    First, we evaluated convergence of phase refinements.
    \Fig{eps/convergence} shows the log \textit{spectral convergence} \cite{merhi11phasereconstruction} of ``PH (4 kHz)'' and ``PH+GD (4 kHz).''
    For comparison, results of randomized initialization of phases are also shown as ``Random.'' 
    The proposed methods (``PH (4 kHz)'' and ``PH+GD (4 kHz)'') have a smaller value for the spectral convergence than ``Random.''
    Also, ``PH+GD (4 kHz)'' has a smaller value than ``PH (4 kHz)'', i.e., ``PH+GD (4 kHz)'' is the closest to the perfect reconstruction. 
    
        \drawfig{t}{0.75\linewidth}{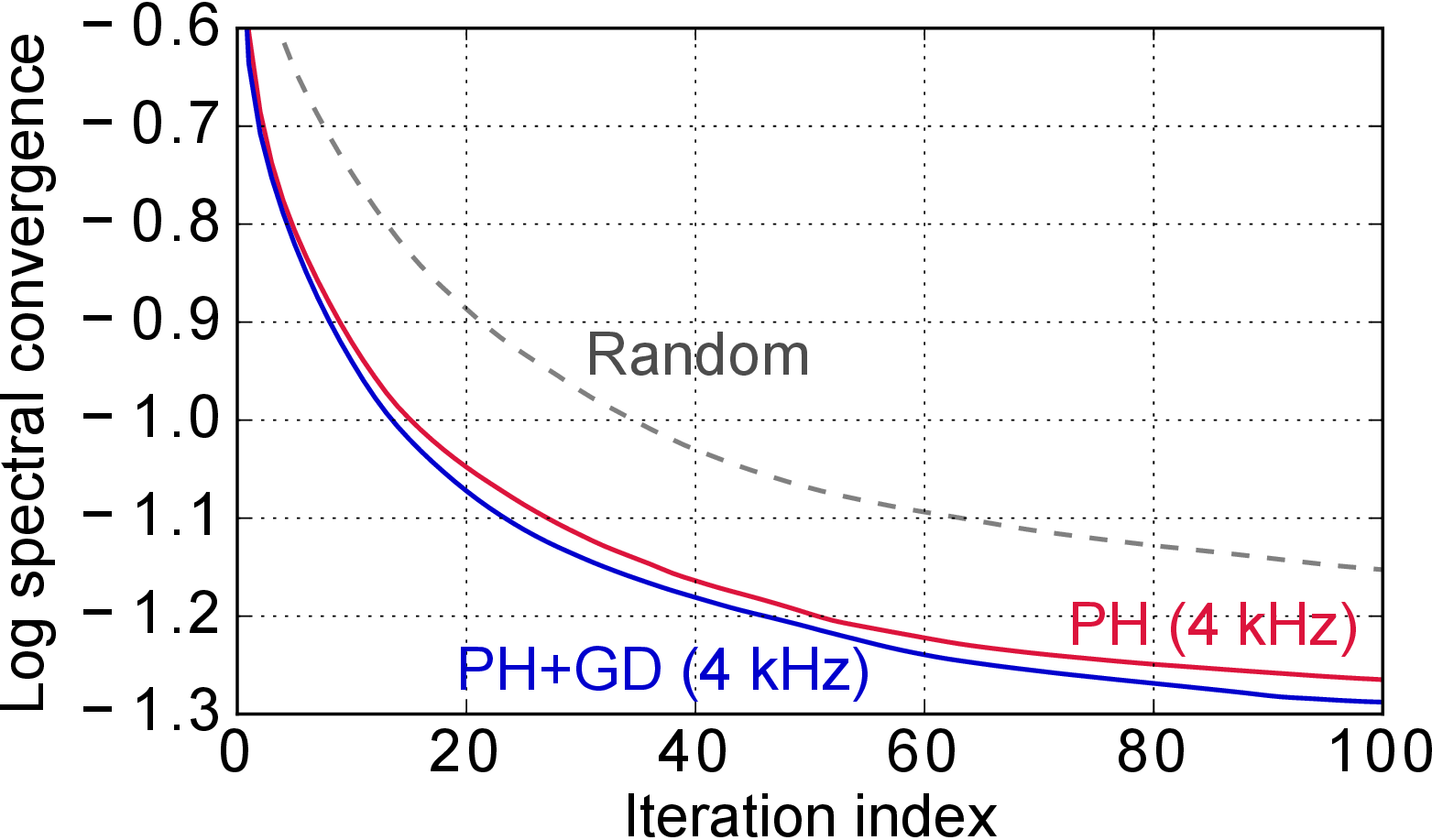}
        {Log spectral convergence by phase refinements. When the value is $-\infty$, perfect reconstruction through STFT and inverse STFT is achieved. 
        This is the result of one of the evaluation datasets, but the same tendency was observed in all evaluation datasets.}

    In addition, we compared speech quality of ``PH'' and ``PH+GD.''
    \Table{subj_group_delay} lists the results of the comparison with frequency bands of 0-2, 0-4, and 0-8 kHz.
    ``PH'' is preferred more in 0-8 kHz with statistical significance, 
    but ``PH+GD'' is preferred more in 0-2 and 0-4 kHz (without statistical significance).
    Therefore, we can demonstrate the effectiveness of group-delay loss in speech quality.

        \begin{table}[t]
        \centering
        \caption{Results of preference tests: effects of group-delay loss.
        \textbf{Bold} indicates preferred method that has a $p$-value smaller than $0.05$}
        \label{tb:subj_group_delay}
        \begin{tabular}{|r|cc|l|}
        \hline Method A & Scores & $p$-value & Method B \\
        \hline   
        \hline PH (2 kHz) & 0.487 vs. 0.513 & 0.514 & PH+GD (2 kHz) \\
        \hline PH (4 kHz) & 0.486 vs. 0.514 & 0.500 & PH+GD (4 kHz) \\      
        \hline PH (8 kHz) & \textbf{0.545} vs. 0.455 & 0.031 & PH+GD (8 kHz) \\       
        \hline
        \end{tabular}
        \end{table}

\Section{Conclusion}
This paper presented DNN-based phase reconstruction from an amplitude spectrogram.
Based on maximum likelihood estimation of the von Mises distribution, we introduced two loss functions for DNN training: phase loss and group-delay loss.
We demonstrated 
1) the trained DNNs can predict group delay more accurately than phases, and 
2) our proposed phase reconstruction methods achieve better speech quality than the conventional Griffin-Lim phase reconstruction method.
For future work, we will 
implement other probability distributions for periodic variables, 
propose other approaches for phase refinements, 
and integrate our method with text-to-speech synthesis that generates amplitude spectrograms \cite{saito18fftssgan}.

\textbf{Acknowledgments:}
Part of this work was supported by SECOM Science and Technology Foundation, and JSPS KAKENHI Grant Number 18K18100.

\bibliographystyle{IEEEbib}
\bibliography{tts}

\end{document}